\documentclass[10pt,preprint]{emulateapj}

\def\msun{\rm M_{\sun}}

\def\micron{$\mu$m}

\begin{document}

\shortauthors{Espaillat et al.}
\shorttitle{Gapped disk around LkCa 15}

\title{Confirmation of a gapped primordial disk around LkCa 15}

\author{Catherine Espaillat\altaffilmark{1}, Nuria Calvet\altaffilmark{1}, Kevin L. Luhman\altaffilmark{2}, James Muzerolle\altaffilmark{3}, and Paola D'Alessio\altaffilmark{4}}

\altaffiltext{1}{Department of Astronomy, University of Michigan, 830 Dennison Building, 500 Church Street, Ann Arbor, MI 48109, USA; ccespa@umich.edu, ncalvet@umich.edu}
\altaffiltext{2}{Department of Astronomy and Astrophysics, The Pennsylvania State University, University Park, PA 16802, USA; kluhman@astro.psu.edu}
\altaffiltext{3}{Steward Observatory, University of Arizona, Tucson, AZ 85712, USA; jamesm@as.arizona.edu}
\altaffiltext{4}{Centro de Radioastronom\'{i}a y Astrof\'{i}sica, Universidad Nacional Aut\'{o}noma de M\'{e}xico, 58089 Morelia, Michoac\'{a}n, M\'{e}xico; p.dalessio@astrosmo.unam.mx}

\begin{abstract}

Recently, analysis of near-infrared broad-band photometry and {\it Spitzer} IRS spectra has led to the identification of a new ``pre-transitional disk" class whose members have an inner optically thick disk separated from an outer optically thick disk by an optically thin gap.  This is in contrast to the ``transitional disks" which have inner disk holes (i.e. large reductions of small dust from the star out to an outer optically thick wall).
In LkCa 15, one of these proposed pre-transitional disks, detailed modeling showed that although the near-infrared fluxes could be understood in terms of optically thick material at the dust sublimation radius, an alternative model of emission from optically thin dust over a wide range of radii could explain the observations as well.  To unveil the true nature of LkCa 15's inner disk we obtained a medium-resolution near-infrared spectrum spanning the wavelength range 2-5 {\micron} using SpeX at the NASA Infrared Telescope Facility.
We report that the excess near-infrared emission
above the photosphere of LkCa 15 is a black-body continuum which can only be due to optically thick material in an inner disk around the star.
When this confirmation of a primordial inner disk is combined with earlier observations of an inner edge to LkCa 15's outer disk it reveals a gapped structure.
Forming planets emerge as the most likely mechanism for clearing the gap we detect in this evolving disk.

\end{abstract}


\section{Introduction}

The origin of a star and its planets is intricately tied to the evolution of the
system's primordial accretion disk.  These disks are composed of gas and dust
and are formed in the collapse of the star's natal, molecular cloud \citep{terebey84}.  As time
passes, the dust grains in these primordial disks evolve:  they collide and
stick, eventually growing in size and perhaps forming planetary systems much
like our own \citep{weiden97}.  The finer details of how the disk material evolves from an
initially well-mixed distribution of gas and dust to a system composed mostly of
large solids like our own solar system is not well understood.  

In recent years
a growing number of primordial disks with signatures of dust evolution hinting
to the early stages of planet formation have been identified.  Most of these
cases, dubbed transitional disks, \citep{strom89} consist of stars with inner holes in their
disks that are mostly devoid of material.  
Within the past few years,
observations at mid-infrared wavelengths by the {\it Spitzer Space Telescope} have led
us to define transitional disks as those objects with small or negligible
near-infrared flux excesses over photospheric fluxes but with a substantial
excess in the mid-infrared and beyond.  This flux deficit at near-infrared
wavelengths relative to full disks accreting material onto the star has been
explained by modeling transitional disks as optically thick disks with inner
cleared regions; the mid-infrared emission originates in the inner edge or
``wall" of the truncated disk which is frontally illuminated by the star \citep{calvet05}.   A
small number of these transitional disks with detailed {\it Spitzer} IRS spectra have
been analyzed to date.  The estimated truncation radii of these disks cover a
wide range, from 4 AU in DM Tau to 24 AU in GM Aur \citep{calvet05}.  Transitional disks have
been found in all ages where protoplanetary disks have been identified, from the
 $\sim$1--2 Myr old Taurus population \citep{calvet05} to the 10 Myr old TW Hya association \citep{calvet02} and 25 Ori (Espaillat et al, in preparation).  In each
case, the disk is accreting mass onto the star so we can conclude that gas is
being transported through the inner cleared disk.  In addition, in some cases a
small amount of micron or sub-micron dust coexists with the gas in this region,
giving rise to an excess over photospheric fluxes detected in the near infrared.  Details of the distribution of this optically thin dust are largely unknown, although near-infrared interferometric observations suggest that this material is highly structured \citep{ratzka07}.

A new class of evolving disks has been identified very recently. Disks in this
class have inwardly truncated outer disks, as the transitional disks.  However,
their significantly larger near-infrared excess, comparable to that of full disks, points to
the existence of a remaining optically thick disk separated by a gap from the
outer disk. A handful of these disks have been analyzed to date, including four
around intermediate mass stars \citep{brown07} and two around classical T Tauri stars \citep{espaillat07b}.  In two of
these cases, LkCa 15 and LkH${\alpha}$ 330, millimeter interferometry has confirmed the truncation of the outer
disk \citep{pietu06,brown08} in agreement with the SED modeling \citep{espaillat07b,brown07}. However, 
the substantial near-infrared excess of LkCa 15 could be
explained by either optically thick material or by 5$\times$10$^{-11}$ ${\msun}$ of optically thin
dust mixed with the gas in the inner disk \citep{espaillat07b}.
These contradictory hypotheses cannot not be properly tested with the existing
near-infrared data, which consists of 2MASS and Spitzer broad-band photometry.

Here, we present detailed spectroscopic data that allows us to unambiguously
search for optically thick material in the inner disk of LkCa 15, a CTTS in the
Taurus cloud, in which an outer disk truncated at 46 AU has been imaged in the
millimeter \citep{pietu06}. 
We confirm
that LkCa 15 belongs to the new class of disks around young stellar objects, the
``pre-transitional disks," \citep{espaillat07b} where a gap is opening within the disk around a low
mass pre-main sequence star.

\section{Observations \& Data Reduction}
\label{sec:obs}
We obtained a 2--5 {\micron} spectrum of LkCa 15 with SpeX at the NASA Infrared Telescope Facility (IRTF) \citep{rayner98}  on December 3, 2007. 
The spectrum was reduced with Spextool  \citep{cushing04} and dereddened with the Mathis
dereddening law \citep{mathis90} and an A$_{V}$ of 1.2 \citep{espaillat07b}. For our template spectrum we use HD36003 (K5 V) which
corresponds to the spectral type of LkCa 15 \citep{kh95} and was obtained from the IRTF
Spectral Library. \footnote
{http://irtfweb.ifa.hawaii.edu/$\sim$spex/spexlibrary/IRTFlibrary.html} 
In Figure 1 we present the medium resolution near-infrared spectra of LkCa 15 and the K5 dwarf template. 

\section{Analysis}

LkCa 15's spectrum has
absorption lines that are weaker than those seen in the spectrum of a
standard star of the same spectral type (Figure 2).  This ``veiling" of the absorption lines is due to an excess continuum that adds to the intrinsic photospheric flux, decreasing the depths of the absorption lines \citep{hartigan89}.
We see this veiling phenomenon in
similar spectra of full primordial disks \citep{muzerolle03} in which it is due to blackbody
emission from the inner disk's optically thick wall located at the radius where
dust sublimates. 
We derived
the veiling (r=F$_{excess}$/F$_{*}$) \citep{hartigan89} by adding an artificial excess continuum to the
template spectrum until the photospheric line depths matched those seen in LkCa
15's spectrum in the K-band.  We measure a veiling of 0.3 at $\sim$2.3 {\micron}, which is consistent
with the excess above the photosphere inferred from broadband photometry \citep{espaillat07b}. The
veiling seen in LkCa 15 cannot be produced by a low-mass companion because such
an object would cause the spectral lines in the composite spectrum of the system
to be stronger than the lines expected from the optical spectral type, rather
than weaker as observed.  

To extract the spectrum of the excess emission (Figure 3), we
follow \citet{muzerolle03} and scale the entire LkCa 15 spectrum according to the above
K-band veiling estimate (Figure 1) and subtract the original template spectrum from it. 
The near-infrared excess emission above the photospheric flux
that is seen in LkCa 15 is well-matched by a single-temperature blackbody of
1600 K (Figure 3), which lies within the range of dust sublimation temperatures
found for a large sample of classical T Tauri stars and Herbig Ae/Be stars \citep{monnier02}.
From this we conclude that the near-infrared excess of LkCa 15 originates from
the wall of an optically thick inner disk located at the dust destruction
radius.  When these results are interpreted with previous Spitzer IRS and
millimeter observations, we can firmly state that this object has a gapped disk
structure (Figure 4), making LkCa 15 a member of the pre-transitional disk class.

\begin{figure}
\figurenum{1}
\epsscale{1.0}
\plotone{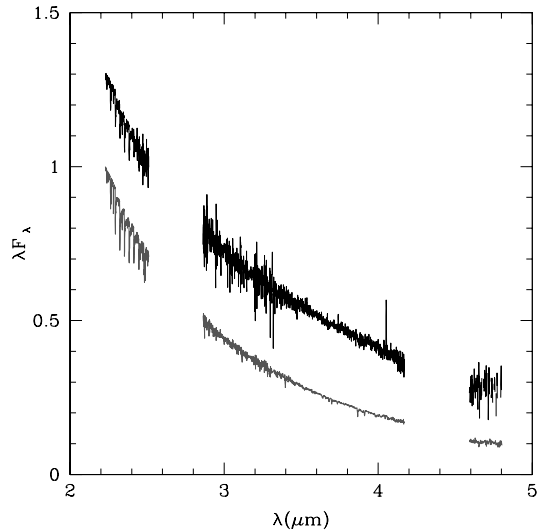}
\caption{
Near-infrared SpeX spectra of LkCa 15 (upper dark line) and a K5 dwarf template (lower light line). Fluxes are in units of the template's flux at $\sim$2.2 {\micron}.  LkCa 15's spectrum is scaled relative to the template by the derived veiling value of 0.3.   Telluric absorption is too strong at 2.5--2.8 {\micron} and 4.2--4.6 {\micron} for useful measurements at these wavelengths.
}
\end{figure}

\begin{figure}
\figurenum{2}
\epsscale{1.0}
\plotone{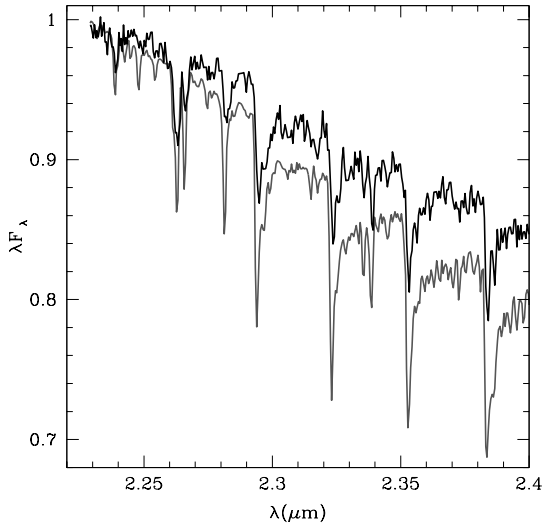}
\caption{
The K-band portion of the spectra in Figure 1.  The LkCa 15 spectrum (upper dark line) has been scaled down to the template's flux at $\sim$2.2 {\micron} in order to more clearly show the veiled absorption lines of LkCa 15 relative to the K5 dwarf template (lower light line).
}
\end{figure}
 
\begin{figure}
\figurenum{3}
\epsscale{1.0}
\plotone{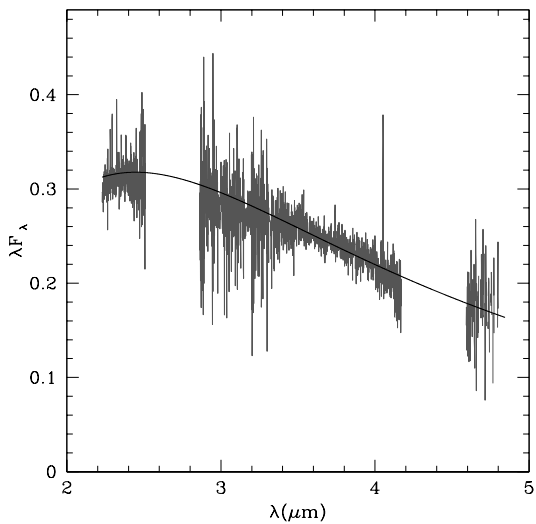}
\caption{
The near-infrared excess spectrum of LkCa 15 (light line) fitted with a 1600 K blackbody (dark line).   Fluxes are in the same units as Figure 1.  The excess was obtained by subtracting the original template spectrum from the veiling-scaled LkCa 15 spectrum which are both shown in Figure 1.
}
\end{figure}

\begin{figure}
\figurenum{4}
\epsscale{1.0}
\plotone{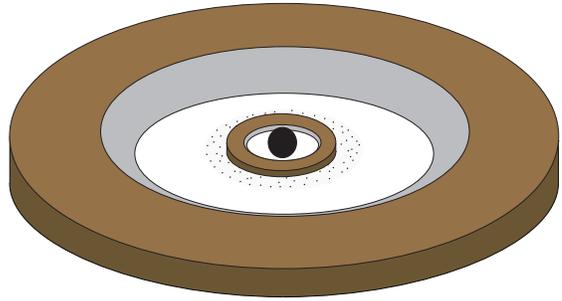}
\caption{
A schematic of the dust distribution in LkCa 15 (not to scale).  The
central circle is the star.  Progressing outward the components of the disk consist of the following: the inner wall
at the dust destruction radius (light gray), the inner disk (brown), a
disk gap (white) with a small amount of optically thin dust (dots), the
outer disk wall (light gray), and the outer disk (brown).  Based on
\citet{pietu06} and \citet{espaillat07b}, LkCa 15's optically thick inner disk is located between
0.12 AU and 0.15 AU and the optically thin dust extends out to 5 AU. 
Between 5 and 46 AU the disk is relatively clear of small dust grains
and the outer disk is inwardly truncated at 46 AU.  
The spectrum of LkCa 15 shown in Figure 1 arises from both the star and inner wall shown in this schematic; the excess emission in LkCa 15 shown in Figure 3 originates only from the inner wall. [See the electronic version of the journal for a color version.] 
}
\end{figure}
   
\section{Discussion \& Conclusions}

Unseen planets \citep{quillen04,rice03} have been proposed to clear out the large cavities within the
primordial disks around GM Aur \citep{calvet05}, DM Tau \citep{calvet05}, TW Hya \citep{calvet02,uchida04,hughes07}, and CoKu Tau/4 \citep{dalessio05}. 
However, several other explanations are also possible.  The inner disk holes in
most of these ``transitional disks" can be explained by inside-out evacuation
mechanisms like the magneto-rotational instability (MRI; Chiang \& Murray-Clay 2007) and, in the case of
CoKu Tau/4, by photoevaporation \citep{alexander07}.  In addition, a stellar companion can
inwardly truncate the outer disk as is most likely the cause of the inner hole
of CoKu Tau/4 \citep{ireland08}.  

\citet{alexander07} found that the accretion rates and disk masses of GM Aur, DM Tau, and TW Hya suggest planet formation in these systems, but 
perhaps the best observational evidence that links planets
with the holes seen in transitional disks is the detection of a planet around TW
Hya \citep{setiawan08}.  Nonetheless, this still does not discount that the MRI is the main clearing agent in the inner disk given that this
mechanism is still viable in the presence of a planet.  Consequently, inner
disk holes are not conclusive signatures of planet formation.  

The origin of the gap in LkCa 15 can be evaluated against different mechanisms
that have been proposed to clear the inner disk, namely the MRI,
photoevaporation, stellar companions, and planet formation.  The MRI operates on
the ionized, frontally illuminated wall of the inner disk and allows material to
accrete onto the star leading to inside-out clearing \citep{chiang07}; the MRI cannot account
for a remnant optically thick inner disk.  In the photoevaporation model, a stellar wind creates
a small gap and halts the inward accretion of mass.  Without replenishment the
inner disk quickly accretes onto the central star and only then can the hole
grow outward \citep{clarke01}. When the hole is about 46 AU, as is seen in LkCa 15, no inner
disk remains \citep{alexander07} so LkCa 15's gap cannot be due to photoevaporation.  A companion
star would have to be located at about 18 -- 26 AU \citep{artymowicz94} in order to truncate the
outer disk at 46 AU and studies of LkCa 15 have revealed no stellar mass
companion down to about 4 AU \citep{leinert93, ireland08}.  Planet formation emerges as the most likely
explanation since a planet can create a gap about its orbit \citep{paardekooper04, varniere06}. The large gap of
LkCa 15, which encompasses the orbits of Mercury (.4 AU) and Neptune (30 AU),
raises the interesting possibility that we are seeing clearing due to multiple planets which would suggest that LkCa 15 may be an
early analog of our own Solar system.

In conclusion, our observations confirm the presence of an inner optically thick disk in LkCa
15. The
existence of optically thick material inside a truncated disk provides
significant insight into the models presented to date to explain the
transitional disks and calls for more detailed studies of this new class of disk.

\vskip -0.1in \acknowledgments{These observations were obtained using SpeX at the Infrared Telescope Facility, which is operated by the University of Hawaii under Cooperative Agreement no. NNX08AE38A with the National Aeronautics and Space Administration, Science Mission Directorate, Planetary Astronomy Program.  We thank Lee Hartmann for insightful discussions.  N.C. acknowledges support from NASA
Origins Grant NNG05GI26G.  K. L. was supported by grant AST-0544588 from the National Science Foundation.  P.D. acknowledges grants from
CONACyT, M\`exico.}

\end{document}